\documentclass[prb,twocolumn,floatfix,showpacs,amssymb]{revtex4}

\usepackage{graphicx}\usepackage{dcolumn}
\usepackage{bm}
\begin{document}

\title{Nonlocal transport mediated by nonlocal Hikami boxes: Condensation
of evanescent quasiparticles injected into the superconducting gap}
\author{S. Duhot}
\affiliation{
Institut NEEL, CNRS \& Universit\'e Joseph Fourier, BP 166,
F-38042 Grenoble Cedex 9}
\author{R. M\'elin}
\affiliation{
Institut NEEL, CNRS \& Universit\'e Joseph Fourier, BP 166,
F-38042 Grenoble Cedex 9}

\begin{abstract}
Evanescent quasiparticles entering a superconductor 
and propagating 
over distances larger than the coherence length give rise
to intermediate states induced by 
double Andreev scattering on disorder.
The resulting effective attractive
interaction between evanescent quasiparticles is retarded
at the extremely slow frequency of the applied bias voltage.
The out-of-equilibrium mesoscopic
superconductor with a fluctuating phase variable is
compatible with a recent experiment
[S. Russo {\it et al.},
 Phys. Rev. Lett. {\bf 95},  027002 (2005)]. Microscopic
theory is discussed in the random phase approximation.
\end{abstract}

\maketitle

\section{Introduction}
The developments 
in nanotechnology allow to  envision the realization
of a source of entangled pairs of electrons \cite{Choi,Martin}, 
the electronic counterpart of a source of Einstein,
Podolsky Rosen pairs of photons, with possible
applications to quantum information and to a test of
non locality of
quantum mechanics with electrons.
Recent
experiments on three terminal devices
by Beckmann {\it et al.} \cite{Beckmann},
Russo {\it et al.} \cite{Russo} 
and more recently by Cadden-Zimansky and Chandrasekhar \cite{Chandra}
realize
an important step in this direction
by probing crossed Andreev
reflection~\cite{Byers,Deutscher,Falci,Samuelson,Prada,Koltai,Feinberg-des,japs,Melin-Feinberg-PRB,Melin-PRB,Levy,Duhot-Melin,Morten,Giazotto,Golubov},
in which Cooper pairs from a superconductor
give rise to 
pairs of correlated electrons in two different ferromagnetic
or normal
electrodes.

\begin{figure}
\includegraphics [width=.8 \linewidth]{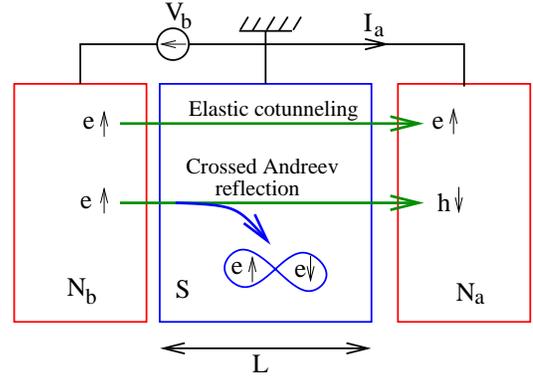}
\caption{(Color online.) Schematic representation of the
electrical circuit in a crossed transport experiment, and of
the two possibilities for crossed transport
in a N$_a$ISIN$_b$ trilayer:
transmission in the electron-electron channel (elastic cotunneling, EC)
and in the electron-hole channel,
leaving a pair in the superconductor (crossed Andreev reflection, CAR).
``e $\uparrow$'' corresponds to a spin-up electron and
``h $\downarrow$'' to a hole in the spin-down band.
The CAR and EC transmission coefficient decay exponentially 
over the coherence length as 
the thickness of the superconductor is increased.
\label{fig:FIG1}
}
\end{figure}

\begin{figure}
\includegraphics [width=.8 \linewidth]{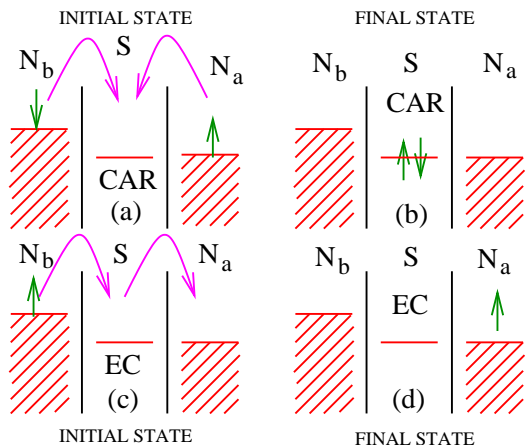}
\caption{(Color online.) Schematic 
representation of the lowest order processes:
crossed Andreev reflection (CAR, (a) and (b))
and elastic cotunneling (EC, (c) and (d)).
Not only has the current 
through electrode ``a'' an opposite sign for CAR ((a) and (b))
and EC ((c) and (d))
(as seen from the arrows on the figure), but the CAR current
is exactly opposite to the EC current for the lowest order
processes in the tunnel amplitudes (as seen from Ref.~\onlinecite{Falci}).
\label{fig:FIG2}
}
\end{figure}

Andreev reflection \cite{Andreev}, the 
mechanism by which an electron
from a normal electrode is reflected as a hole at a normal
metal-superconductor (NS) interface 
while a pair is transmitted into the superconductor,
takes place in a region of size $\xi$,
the coherence length of the disordered superconductor.
Crossed Andreev reflection
(CAR) in a three terminal
normal metal-insulator-superconductor-insulator-normal metal
(N$_a$ISIN$_b$) trilayer
corresponds to an Andreev process
such that an electron from N$_b$ is transmitted as a hole in N$_a$
over a distance of order $\xi$,
leaving a pair in the superconductor.
Elastic cotunneling (EC), the other competing channel,
amounts to transporting an electron from
N$_b$ to N$_a$ without changing its spin. 
The two possibilities are
shown schematically on Figs.~\ref{fig:FIG1} and~\ref{fig:FIG2}.
The nature of the dominant crossed transport channel can be controlled
by the relative spin orientation of strongly polarized
ferromagnets in a 
ferromagnet-insulator-superconductor-insulator-ferromagnet (F$_a$ISIF$_b$)
structure \cite{Deutscher,Falci}:
EC (CAR) dominates 
with (anti)parallel spin orientations, as it can be seen from
the spin of the electron or hole transmitted in electrode ``a''
(see Fig.~\ref{fig:FIG1}).
On the other hand, EC dominates for normal metals 
with highly transparent interfaces \cite{Melin-Feinberg-PRB}.
Crossed transport dominated either by 
EC or by CAR is obtained in the three recent experiments 
mentioned above
\cite{Beckmann,Russo,Chandra}.

More precisely, the crossed conductance \cite{Lambert,Jedema}
${\cal G}_{a,b}(V_b)$ of a three terminal device
measures
the sensitivity of the
current $I_a(V_a,V_b)$ flowing through electrode ``a'' with voltage $V_a$
to the voltage $V_b$ applied on  electrode ``b'': 
\begin{equation}
\label{eq:Gab}
{\cal G}_{a,b}(V_a,V_b)=\frac{\partial I_a(V_a,V_b)}
{\partial V_b}
\end{equation} 
(see the circuit on Fig.~\ref{fig1}a).
The voltage $V_a$ may be arbitrary but, according to available experiments
\cite{Russo,Beckmann,Chandra},
we consider that electrode ``a'' is grounded ($V_a=0$), as the
superconductor.

Theoretically, the crossed conductance defined by Eq.~(\ref{eq:Gab})
vanishes to lowest
order in the tunnel amplitudes for a N$_a$ISIN$_b$ trilayer
because
the electron and hole transmitted by EC and CAR 
have opposite charges,
and because of a symmetry of crossed transport: CAR
and EC have identical transmission coefficients
\cite{Falci,Feinberg-des,Melin-Feinberg-PRB} (see Fig.~\ref{fig:FIG2}).
This property does not hold for a single channel
ballistic superconductor connected by single atom contacts to normal
electrodes: in this case, the crossed conductance to lowest
order in the tunnel amplitudes oscillates
around zero with periodicity $\lambda_F$ (the Fermi wave-length)
as the distance $L$ between the contacts is varied.
Oscillations in the Fermi phase factors average to
zero in the realistic case of a multichannel contact \cite{Falci},
or in the case of a diffusive superconductor \cite{Feinberg-des}
with non magnetic impurities.

The theoretical prediction of a vanishing of the crossed conductance in a
N$_a$ISIN$_b$ trilayer is however contradicted by the recent experiment 
by Russo {\it et al.} \cite{Russo}, that provides in addition
evidence of a disappearance of the crossed signal in 
a magnetic field applied parallel to the layers. 
A 
characteristic energy scale $\hbar\omega_c$ within the superconducting gap
$\Delta$
is found experimentally~\cite{Russo},
at which the crossed signal changes sign from EC to CAR as the
bias voltage energy increases above 
$\hbar \omega_c$, and eventually vanishes at higher energies.
The experimental $\hbar\omega_c$ decreases to zero as the superconductor
thickness increases.
One may conjecture that such experiments are characteristic
of a geometry-dependent 
coupling of the superconducting phase variable to crossed
transport by evanescent waves.
As we show, a natural possibility
for such coupling (quantum interference effects in the superconductor)
results in an energy scale $\hbar \omega_c$ smaller than
the superconducting gap for extended junctions.
Another possibility (Coulomb interactions)
was put forward by Levy Yeyati {\it et al.}
in Ref.~\onlinecite{Levy}, in connection with
the modes of the electromagnetic environment.

\section{Quantum interference effect in superconductors}

\subsection{Transmission modes related to evanescent waves}
Weak localization enhances the return probability and the phase
fluctuations of
a normal metal \cite{Mon}. A superconducting condensate
is on the contrary delocalized and has a well defined phase.
Superconductivity and weak localization are related to each other
because
the Cooper pairs of a superconductor can participate to the Cooperons
of weak localization. 
Smith and Ambegaokar show indeed
that the phase stiffness of a superconductor is reduced
by weak localization \cite{Smith}.
Weak localization
is however strongly modified in subgap transport through
a superconductor by the following property of
evanescent wave functions.

Electron tunneling through a normal metal involves the
simultaneous forward propagation in time of an electron and the
backward propagation in time of a 
hole, forming a diffuson. 
The terminology ``transmission
mode'' is used here
for describing the effect of disorder on subgap tunneling.
For evanescent states, a hole
propagating backward in time can be replaced in a
transmission mode by an electron propagating forward in time:
the wave-function of a normal
electron in a box of size $\xi$ being real valued, is equal to
the time reversed particle wave-function
in the absence of a magnetic field. 
Transmission modes of range $\xi$ in subgap transport
through a superconductor can thus  
be made also either of a pair of electrons or of a pair
of holes propagating forward in time
\cite{Altland}. 
In a superconductor,  a
pair of electrons propagating forward in time  
can be obtained from a Cooper pair in the condensate.

\subsection{Electron-hole conversion in transmission modes}
\begin{figure}
\includegraphics [width=.8 \linewidth]{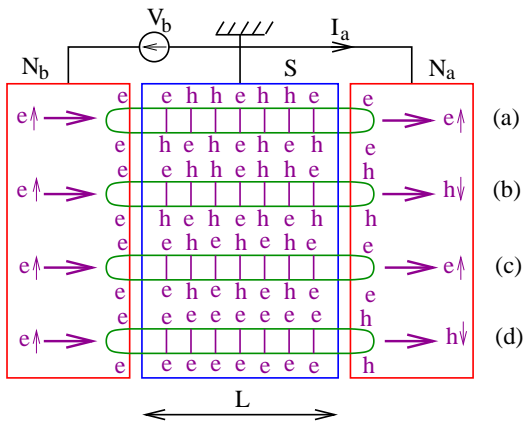}
\caption{(Color online.) Schematic representation of the ladder
series capturing the influence of disorder on the processes on
Fig.~\ref{fig:FIG1}.
Examples of Nambu labels (``e'' for electron and ``h'' for hole)
are indicated on the figure,
for elastic cotunneling with transmission of an electron
((a) and (c))
and for crossed Andreev reflection with transmission of a hole 
((b) and (d)).
``e $\uparrow$'' corresponds to a spin-up electron and
``h $\downarrow$'' to a hole in the spin-down band. For clarity, the 
$\uparrow$ and $\downarrow$ symbols of electrons and hole are
not shown for the diffuson.
\label{fig:basic-diff}
}
\end{figure}

As discussed in the preceding section,
transmission modes in a disordered superconductor correspond to
two quasiparticles
scattering on the same sequence of impurities. Transmission modes
are described 
in a standard way
on the basis of ladder diagrams (see for instance
Smith and Ambegaokar \cite{Smith} in the context of localization 
in a superconductor).
As a direct consequence of Ref.~\onlinecite{Smith},
intermediate states with
electron-hole conversion from one impurity to the next
such as on Fig.~\ref{fig:basic-diff}c
appear as the result of
{\it disorder scattering}, as opposed to 
usual Andreev reflection at a normal metal-superconductor interface
\cite{BTK}
being induced by {\it spatial
variations of the superconducting pair potential}.
Intermediate states with
electron-hole conversion in the disordered case
are of course not due to
anomalous electron-hole scattering on impurities, but to 
electron-hole conversion during propagation
in between two impurities.

For instance,
transmission of an electron from electrode N$_b$ as an
electron in electrode N$_a$, and transmission of an electron
in electrode N$_b$ as a hole in electrode N$_a$ are shown on
Figs.~\ref{fig:basic-diff}a and b respectively.
The extreme case of
a sequence of electron-hole conversions due to disorder,
resulting in
net transmission from electrode N$_b$ to electrode N$_a$
in the electron-electron channel
is shown on Fig.~\ref{fig:basic-diff}c. Another case of
net transmission of an electron in electrode
N$_b$ as a hole in electrode N$_a$
by a sequence of normal scattering
on disorder is shown on Fig.~\ref{fig:basic-diff}d.
Electron-hole conversion in between two impurities on
Fig.~\ref{fig:basic-diff}c corresponds to a usual Andreev
process, however ``interrupted'' at the time scale of the
elastic scattering time,
leading to intermediate virtual states with a characteristic energy
much above the superconducting gap.
We show below that this type of intermediate state
characteristic energy scale
for Andreev processes ``internal'' to the superconductor
can be very much reduced by quantum interference effects
and become observable.

\subsection{Self-crossings of transmission modes}
\begin{figure}
\includegraphics [width=.8 \linewidth]{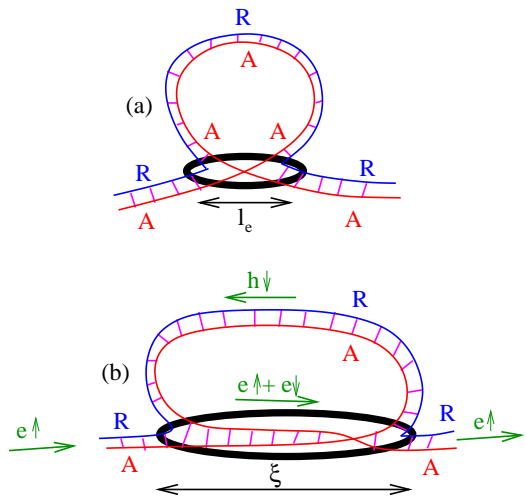}
\caption{(Color online.) Schematic representation of transmission
mode self-crossings in the presence of disorder. The circled
area is the Hikami box\cite{GLK,H}, of dimension set by the elastic
mean free path $l_e$ in a normal metal, and of dimension set
by the coherence length $\xi$ for subgap transport.
\label{fig:self-crossing}
}
\end{figure}

\begin{figure}
\includegraphics [width=.75 \linewidth]{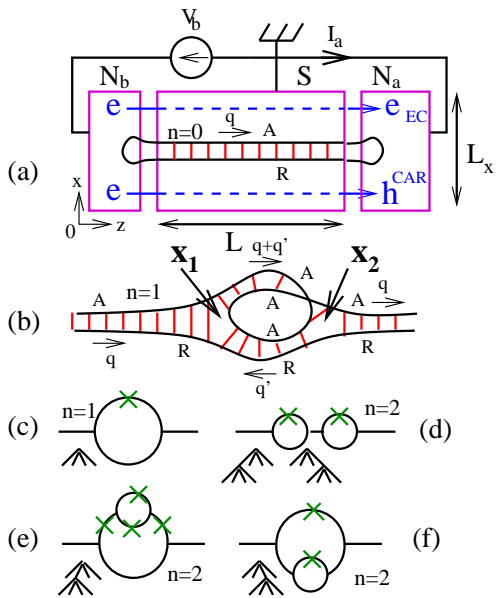}
\caption{(Color online.) Representation of (a) 
the electrical circuit for crossed transport through a 
N$_a$SN$_b$ trilayer, the electron e (top) and hole h 
(bottom) transmitted in electrode ``b'' by EC and CAR,
and the ``bare'' transmission mode
of wave-vector ${\bf q}$ in
the ladder approximation.  The junction has a dimension
$L_y$ along the $y$ axis 
perpendicular to the figure. The aspect ratio is not to
the scale of the experiment \cite{Russo} where $L\simeq 15
\div 200\,$nm, and $L_x,L_y\simeq 4,8\,\mu$m.
(b) is a single weak localization-like loop in a 
transmission mode
self-crossing.
The normal case is recovered 
for an AA transmission modes propagating locally over the elastic mean
free path.
(c) is a compact representation of (b) and of the associated
tree. The four leaves correspond to the four transmission modes
making a weak localization-like loop.
AR (AA) transmission modes
are represented by solid lines (with a cross). 
(d), (e) and~(f) represent
the random phase approximation (RPA) diagrams with $n=2$ loops
and the associated trees.
\label{fig1}
}
\end{figure}

Self-crossings of transmission modes \cite{GLK,H} are 
the building blocks of weak localization in the
normal case \cite{Mon,note-diff} (see Fig.~\ref{fig:self-crossing}a).
The distinguishing features of diffuson self-crossings in
subgap transport by evanescent wave functions (see
Figs.~\ref{fig:self-crossing}b and~\ref{fig1}b)
were already pointed in Ref.~\onlinecite{Duhot-Melin}. The necessity
of accounting for ``advanced-advanced'' or ``retarded-retarded''
transmission modes was already discussed by Altland and
Zirnbauer \cite{Altland} for an Andreev quantum dot.

\begin{figure*}
\includegraphics [width=.7 \linewidth]{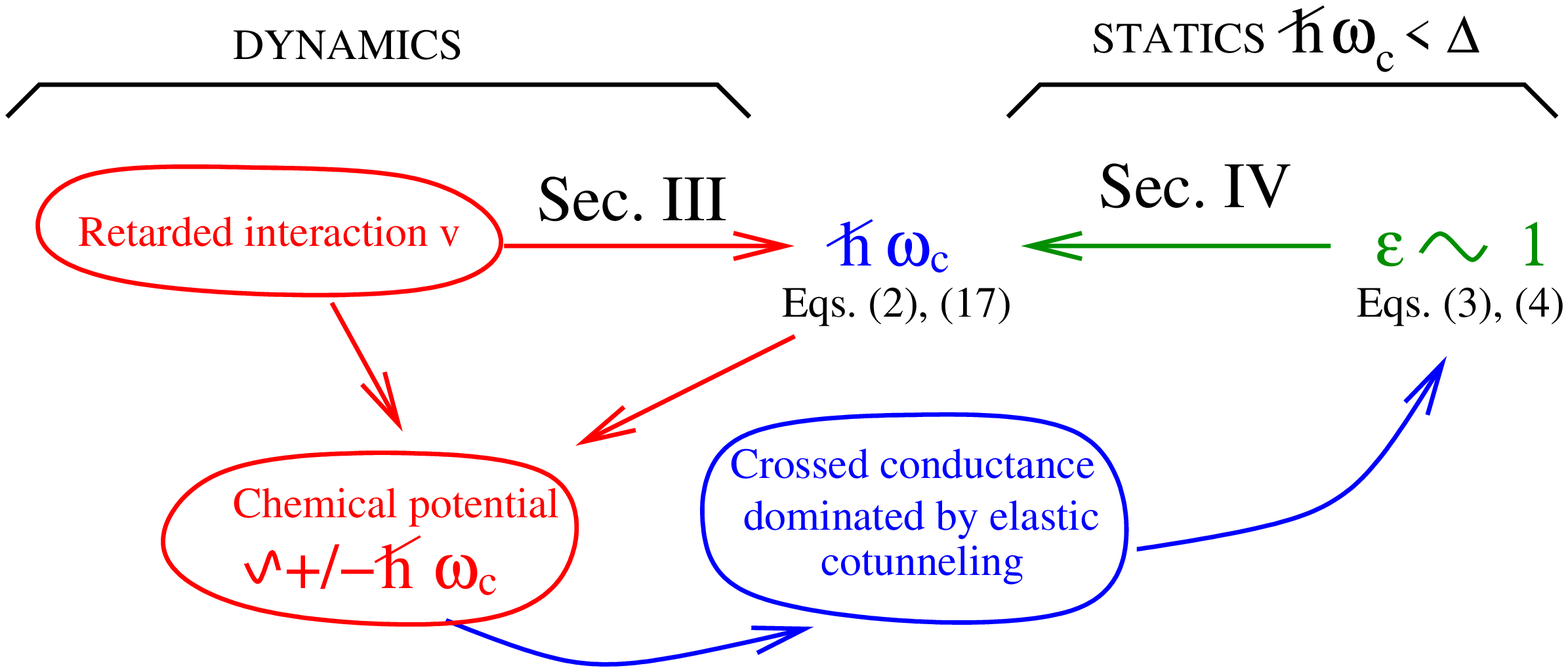}
\caption{(Color online.) Schematic representation the link
between Secs.~\ref{sec:slow} and~\ref{sec:rpa}. 
A characteristic energy scale $\hbar \omega_c$ is obtained from
symmetry breaking between elastic cotunneling and crossed
Andreev reflection ($\epsilon\sim 1$) in Eq.~(\ref{eq:sb-EC})
and (\ref{eq:sb-CAR}). The energy scale $\hbar \omega_c$ is identified to
the characteristic frequency for phase
fluctuations in the presence of a very slow attractive interaction,
which, in turns, favors a symmetry breaking in favor of
elastic cotunneling.
\label{fig:interweaving}
}
\end{figure*}

Weak localization-like loops such as on Figs.~\ref{fig:self-crossing}b
and~\ref{fig1}b
contain, among all possibilities, 
intermediate states corresponding to double Andreev processes 
in which an evanescent electron-like quasiparticle (an AR 
transmission mode
made of an advanced and a retarded Green's function for
an electron propagating forward in time and a hole propagating backward
in time) 
transforms by a first Andreev process at ${\bf x}_1$ in
an evanescent hole-like quasiparticle (another AR
transmission mode for a hole
propagating forward in time and an electron propagating backward
in time)
and the transmission of a pair (an AA transmission mode made of
two advanced Green's function for a pair of electrons
propagating forward in time), that
recombine at ${\bf x}_2$ by another Andreev process
after a propagation over $|{\bf x}_2-{\bf x}_1|\sim\xi$
\cite{Duhot-Melin}. The intervention of the condensate in such
loop processes is obvious from noting that
the contribution of weak localization-like processes to the
crossed conductance couples to a phase gradient in the superconductor
\cite{Duhot-soumis-PRB}.
These processes
result in an attractive interaction of
strength $v$ per carrier injected in the gap,
in units of the Fermi energy $\epsilon_F$.
This interaction due to quantum interference effects
has all the features of a {\it very slow} phonon exchange: it is attractive,
and retarded by a time $2\hbar \pi/eV_b$ (because weak localization-like
processes induces transitions only at energy $\hbar \omega=e V_b$).

\begin{figure}
\includegraphics [width=.75 \linewidth]{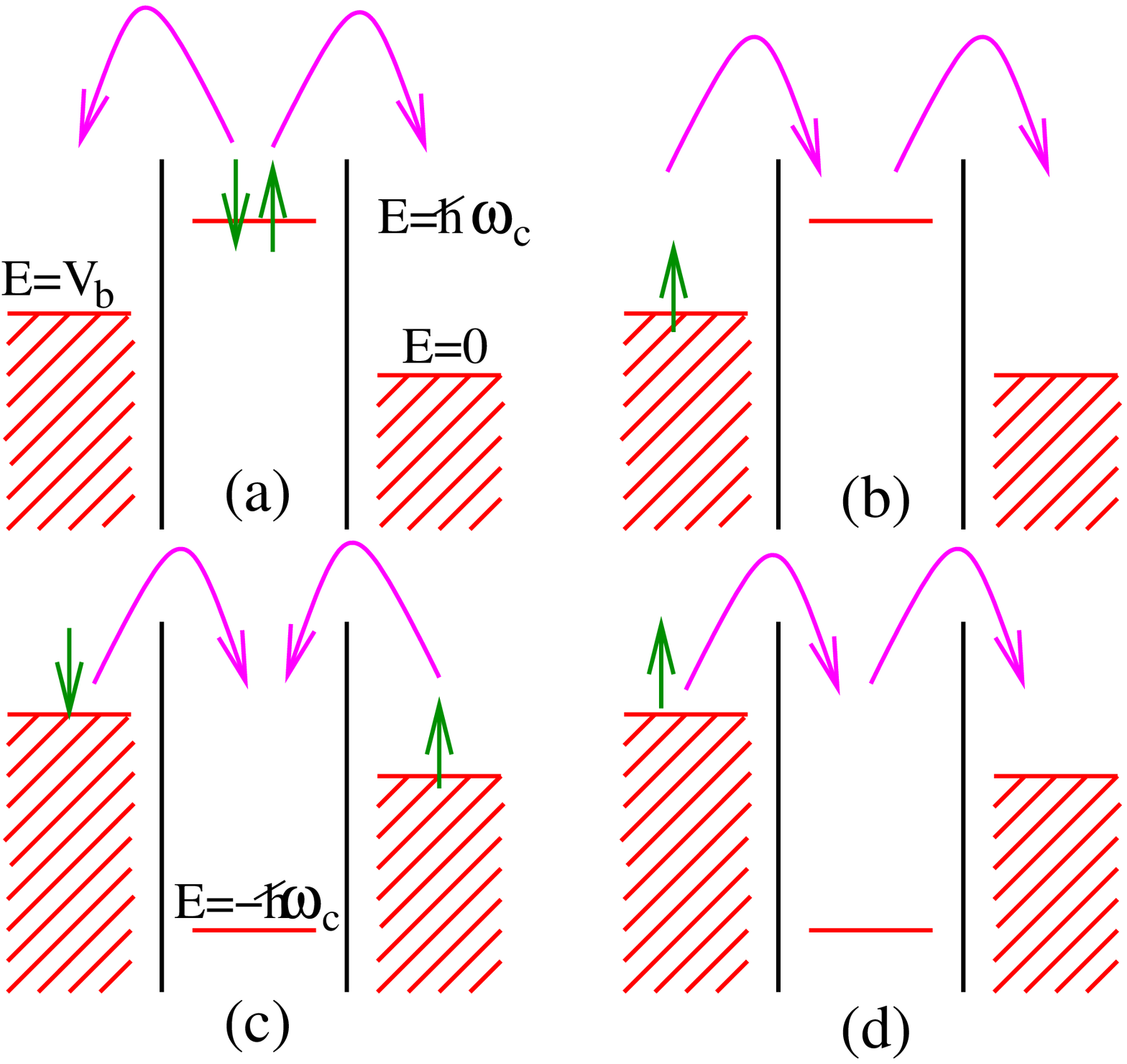}
\caption{(Color online.) Schematic representation the lowest order
processes for superconductor chemical potentials $\hbar \omega_c$
larger than $e V_b$ ((a) and (b)), or smaller than $e V_b$ ((c) and (d)).
The crossed Andreev reflection (CAR)
processes (a) and (c) lead to opposite crossed currents for
superconductor chemical potentials
$\simeq \pm \hbar \omega_c$ while
the elastic cotunneling (EC) processes (b) and (d)
result in additive crossed currents. The crossed conductance
with a superconductor chemical potential fluctuating between
$\simeq \pm \hbar \omega_c$ is thus dominated by EC.
\label{fig:lowest}
}
\end{figure}

As discussed previously, a transmission mode in the ladder
approximation induces electron-hole conversion due to scattering
on disorder, having the elastic scattering time as a characteristic
time/inverse energy scale. 
Weak localization-like loops such as on
Fig.~\ref{fig:self-crossing}b correspond to the transport of
pairs over a distance comparable to the coherence length $\xi$,
and have thus the superconducting gap $\Delta$ as a characteristic
energy\cite{Melin-PRB}. Similar double Andreev processes were discussed in
Ref.~\onlinecite{Duhot-Melin} in a NISIN three-terminal device, and,
interestingly, were introduced previously by Jacobs and
K\"ummel\cite{Jacobs} for the time evolution of wave-packets.
As we show now, even smaller
energy scales are obtained from many interacting
weak localization-like loops. The link between 
an effective dynamical
model for a very slow attractive interaction in addition to the
BCS pairing (Sec.~\ref{sec:slow}) and the resummation of
many weak localization-like loops in the static case
(Sec.~\ref{sec:rpa})
is shown schematically on Fig.~\ref{fig:interweaving}.

\section{Slow attractive interaction in a superconductor}
\label{sec:slow}
\subsection{Energy scale within the gap}
Changing the phase by $\pm 2\pi$ during a time interval
$2\pi/\omega_c$ requires the time $2\pi/\omega_c$ to be smaller than
$2\pi\hbar/eV_b$, the retardation in the interactions that
prevent the phase from fluctuating. The BCS case with a phase
constant in time corresponds to retarded interactions
with a Debye energy $\hbar \omega_D$,
orders of magnitude larger than the superconducting gap
$\Delta$.
In the opposite limit  $e V_b<\hbar \omega_c$ that we consider,
the interaction $v$ has
to visit and to fix the phase of
a number of state $N(0)\hbar \omega_c$ in the normal metal
to rotate the phase by $2\pi$ within a delay $2\pi/\omega_c$,
with $N(0)=2 \pi L L_x L_y/\lambda_F^3 \epsilon_F$ the
number of normal states per unit energy, and with
$\epsilon_F$ the Fermi energy and $\lambda_F$ the
Fermi wave-length
(see $L$, $L_x$ and $L_y$ on Fig.~\ref{fig1}).
By the uncertainty relation $v\epsilon_F \Delta t=\hbar$,
the effective number of interactions in a time interval
$2\pi/\omega_c$ is $N_{int}(\omega_c)=2\pi/\omega_c \Delta t
=2\pi v \epsilon_F/\hbar \omega_c$.
The characteristic energy $\hbar \omega_c$ is
obtained by equating
$N_{int}(\omega_c)=N(0)\omega_c$:
\begin{equation}
\label{eq:omc0}
\frac{\hbar \omega_c}{\epsilon_F}=
\sqrt{\frac{v \lambda_F^3}{LL_xL_y}}
.
\end{equation}
The decoupling between $\hbar \omega_c$ and the states at
energy larger than $\Delta$ is justified if
$\hbar \omega_c \ll \Delta$, which we suppose in the following.
The superconducting phase fluctuates by $\pm 2 \pi$ 
with a frequency $\omega_c$, but its average value over a time
window larger than the retardation of interactions
$2\pi\hbar/e V_b$ does not fluctuate.
The Andreev current from the superconductor to electrode
N$_a$ vanishes for $e V_b<\hbar \omega_c$ because
the superconductor chemical potential
fluctuates between typical values $\pm \hbar \omega_c$.
The crossed conductance is then dominated by EC, 
as it can be seen from 
second order perturbation theory in the tunnel
amplitudes \cite{Spivak} (see Fig.~\ref{fig:lowest}).

\subsection{Qualitative behavior of the crossed conductance}
We consider now the case $e V_b \agt \hbar \omega_c$.
The proximity of the characteristic
energy $\hbar \omega_c$ induces pair correlations
among evanescent
quasiparticles injected in the superconductor
at bias voltage energies
$e V_b \agt \hbar \omega_c$.
CAR and EC processes are described schematically~\cite{Spivak}:
(b), (c) and (d) in the insert of Fig.~\ref{fig2}
correspond to acting twice with the tunnel
Hamiltonian, starting from the superconductor in the
BCS ground state for (b) and
with a correlated pair in the
initial state for
(c) and (d).
The crossed signals due to the EC and CAR processes (c) and
(d) cancel with each other because (c) involves an extra
anticommutation of fermions compared to (d).
The EC process (c) cancels with a CAR process
(not shown on Fig.~\ref{fig2}) in which two electrons
at energies $-eV_b$ and $eV_b$ in electrodes N$_a$
and N$_b$ enter the superconductor as a pair.
Fluctuations by one quasiparticle with spin-$\sigma$ in
the superconductor do not couple to the
crossed signal because
EC has an opposite sign for the two orientations of $\sigma$.
The remaining contribution to the crossed signal
(not shown on Fig.~\ref{fig2})
is due to
tunneling of the two electrons of a correlated pair 
from the superconductor to electrode
N$_a$,
leading to a negative differential crossed resistance,
as for local Andreev reflection above the superconducting gap
in a NIS junction \cite{BTK}.
From these arguments,
we conclude that a change of sign from a positive EC crossed
resistance for $e V_b<\hbar \omega_c$ to a negative 
Andreev reflection
crossed resistance for $e V_b \agt \hbar \omega_c$ occurs,
and that the crossed signal disappears at higher energies,
as in experiments \cite{Russo}.

\section{Random phase approximation 
for weak localization-like processes}
\label{sec:rpa}
\subsection{Method and algorithm}
\subsubsection{Crossed conductance and resistance}
Now, we start from a static superconducting gap
(see Fig.~\ref{fig:interweaving}) and determine
from microscopic theory the analog of Eq.~(\ref{eq:omc0}) 
for a dirty superconductor.
We denote by
$T_{CAR}({\bf R},\omega)$ and
$T_{EC}({\bf R},\omega)$
the CAR and EC dimensionless transmission coefficients
of a superconductor at distance ${\bf R}$ and
energy $\hbar\omega$ below the gap corresponding respectively to 
electron and hole
transmission (see Figs.~\ref{fig:FIG1} and~\ref{fig:FIG2}) described
by transmission modes in the ladder approximation such
as on Fig.~\ref{fig:basic-diff}.

We enforce a finite crossed current by breaking the
symmetry $T_{EC}({\bf R},\omega)=T_{CAR}({\bf R},\omega)$ according to
\begin{eqnarray}
\label{eq:sb-EC}
T_{EC}({\bf R},\omega)&=&T_0({\bf R},\omega)(1+\epsilon)\\
T_{CAR}({\bf R},\omega)&=&T_0({\bf R},\omega)(1-\epsilon)
\label{eq:sb-CAR}
.
\end{eqnarray}
The crossed conductance per conduction channel
\begin{equation}
\label{eq:Gab-R}
{\cal G}_{a,b}({\bf R},\omega)=-2 \left(\frac{e^2}{h}\right) 
{\cal T}^2
T_{ch}({\bf R},\omega)
\end{equation} 
corresponds to processes
of lowest order ${\cal T}^2$
in an expansion in the dimensionless interface
transmission coefficient $0<{\cal T}<1$, and
the charge transmission coefficient of the superconductor
$T_{ch}({\bf R},\omega)=T_{EC}({\bf R},\omega)-T_{CAR}({\bf R},\omega)$
accounts for
the different carriers transmitted by EC and CAR in electrode ``a''.
The resulting crossed resistance per channel is obtained as the
inverse of the extra diagonal element of the inverse of the
crossed conductance matrix:
\begin{eqnarray}
\label{eq:inverse}
\left[\begin{array}{cc}
{\cal R}_{a,a} & {\cal R}_{a,b} \\
{\cal R}_{b,a} & {\cal R}_{b,b} \end{array}\right]
=
\left[\begin{array}{cc}
{\cal G}_{a,a} & {\cal G}_{a,b} \\
{\cal G}_{b,a} & {\cal G}_{b,b} \end{array}\right]^{-1}
,
\end{eqnarray}
with
\begin{equation}
\label{eq:Gaiaj}
{\cal G}_{a_i,a_j}(V_a,V_b)=\frac{\partial I_{a_i}}
{\partial V_{a_j}}(V_a,V_b)
\end{equation}
generalizing Eq.~(\ref{eq:Gab}),
where the entries $a_i$ and $a_j$ in Eq.~(\ref{eq:Gaiaj})
correspond to the normal
electrode labels ``a'' and ``b''.
We deduce from Eq.~(\ref{eq:inverse}) the off-diagonal matrix element
of the crossed resistance matrix:
\begin{equation}
\label{eq:toto1}
{\cal R}_{a,b}=
-\frac{{\cal G}_{a,b}}{{\cal G}_{a,a}
{\cal G}_{b,b}
-{\cal G}_{a,b}{\cal G}_{b,a}}
,
\end{equation}
approximated as
\begin{equation}
\label{eq:toto2}
{\cal R}_{a,b}\simeq -\frac{
{\cal G}_{a,b}}{{\cal G}_{a,a} 
{\cal G}_{b,b}}
,
\end{equation}
because of the damping of crossed processes over the coherence
length $\xi$.
The dependence on $V_a$ and $V_b$ is not explicit in
Eqs.~(\ref{eq:toto1}) and~(\ref{eq:toto2}).
We obtain from Eq.~(\ref{eq:Gab-R}):
\begin{equation}
{\cal R}_{a,b}({\bf R},\omega)\simeq\left(\frac{h}{2e^2}\right)
{\cal T}^{-2}T_{ch}({\bf R},\omega)
,
\end{equation}
much larger than 
for our previous approach for higher order processes at the 
interfaces \cite{Melin-PRB,Duhot-Melin}.

\subsubsection{Algorithm}
The excitations of the condensate, integrated out by 
the random phase approximation (RPA) \cite{Anderson}
for the transmission modes,
lead to a new value of the crossed
conductance corresponding to replacing $\epsilon$ by
$\tilde{\epsilon}({\bf R},\omega,\epsilon)$.
We identify the characteristic energy $\hbar \omega_c$
as the divergence in the energy dependence of
$\tilde{\epsilon}({\bf R},\omega,\epsilon)$,
as for a gap edge singularity.
The symmetry breaking parameter
$\tilde{\epsilon}_n({\bf R},\omega,\epsilon)$ with
$n$ localization loops is obtained by inverting
\begin{eqnarray}
\label{eq:TOTO3}
\tilde{T}_{EC}(n,\epsilon)&=&
\tilde{T}_0(n)+
\tilde{\epsilon}_n(\epsilon) T_0\\
\label{eq:TOTO4}
\tilde{T}_{CAR}(n,\epsilon)&=&
\tilde{T}_0(n)-
\tilde{\epsilon}_n(\epsilon) T_0
,
\end{eqnarray}
with $n$ the number of weak localization-like loops, and where
the dependence on ${\bf R}$ and $\omega$ is not explicit
in Eqs.~(\ref{eq:TOTO3}) and~(\ref{eq:TOTO4}).
We deduce the value of
\begin{equation}
F_n({\bf q},\omega,\epsilon)
\equiv\frac{\tilde{\epsilon}_n({\bf q},\omega,\epsilon)}
{\epsilon}-1,
\end{equation}
where we changed variable from ${\bf R}$ to the wave-vector ${\bf q}$
by a Fourier transform.
\begin{figure}
\includegraphics [width=.9 \linewidth]{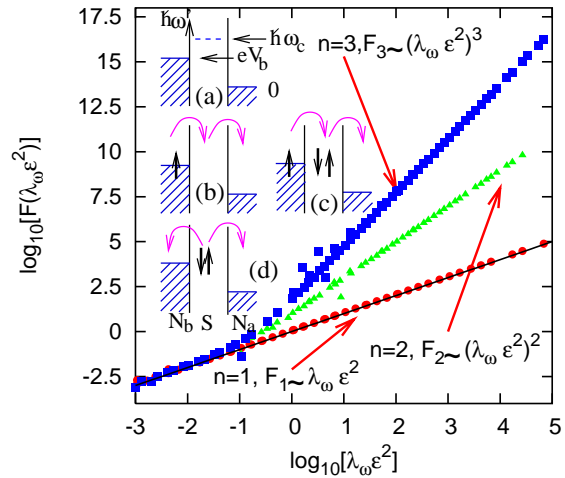}
\caption{(Color online.) Scaling plot of
$F_n(\lambda_\omega \epsilon^2)$ 
as a function of $\lambda_\omega \epsilon^2$ 
for diagrams with $n=1$ ($\bullet$), $n=2$ 
($\blacktriangle$) and $n=3$ ($\blacksquare$)
weak localization-like loops. 
The energy
$\hbar\omega$
and the dimensionless symmetry breaking parameter $\epsilon$
vary over 3 orders of magnitude.
The values of the elastic mean free path $l_e$, such that
$2\le\xi/l_e\le14$ (as compared to
$\xi/l_e=5\div 7$ in the experiment \cite{Russo}),
vary from the dirty limit 
to the cross-over with the ballistic limit.
For clarity, not all calculated points are shown on the figure.
(a) shows one value $2\pi/\omega_c$ of the superconductor
chemical potential.
(b), (c) and (d) show EC and CAR processes for
$e V_b\agt \hbar \omega_c$ with residual pair correlations
among evanescent quasiparticles.
\label{fig2}
}
\end{figure}

An algorithm generates the
topologically inequivalent
higher order RPA diagrams for the transmission mode
with the tree structure shown on
Figs.~\ref{fig1}c, d, e, f.
These diagrams maximizing the number of imbricated loops 
for a given number of branchings are the most
relevant for describing intermediate states with a proliferation of Andreev
reflections internal to the superconductor 
for $\omega \simeq \omega_c$.
In order to reduce the computation time,
we first specialize to a single channel by
restricting to the transverse components
$(q'_x,q'_y,q'_z)=(0,0,2\pi n_z/L)$ of the wave-vector
${\bf q}'=
(q'_x,q'_y,q'_z)=(2\pi n_x/L_x,2\pi n_y/L_y,2\pi n_z/L)$
in the weak localization-like loop,
with $n_x$, $n_y$ and $n_z$ three integers (see 
Fig.~\ref{fig1}).
The scaling with the number of 
longitudinal components is restored afterwards.

\subsection{Results}
Now, we present the result of RPA resummation.
The ${\bf q}$-dependence of the transmission coefficients
is, up to an overall rescaling,
the same as 
for the bare transmission coefficient,
because of the properties of the convolution of 
exponentials in the corresponding transmission coefficients
in real space.
We thus evaluate only the rescaling factor $F_n(0,\omega,\epsilon)$
that
collapses on master
curves (see Fig.~\ref{fig2}) when plotted as a function
of the dimensionless parameter
\begin{equation}
\label{eq:lambda-om}
\lambda_\omega \epsilon^2=k_F L \left(\frac{\hbar\omega
  \epsilon}{\epsilon_F}\right)^2
\left(k_F l_e\right)^{2 \alpha}
,
\end{equation}
according to $F_n(0,\omega,\epsilon)\equiv F_n(\lambda_\omega \epsilon^2)$,
with $l_e$ the elastic mean free path such that $2\le \xi/l_e\le 14$,
and with $\alpha= 1.1\pm 0.1$.
Summing the RPA diagrams leads to $\tilde{\epsilon}$,
the new value of $\epsilon$ modified by the response
of the condensate:
\begin{equation}
\label{eq:epsinf}
\tilde{\epsilon}(\omega,\epsilon)=\epsilon(1+F_\infty(
\lambda_\omega \epsilon^2))=
\frac{\epsilon}{1-\lambda_\omega\epsilon^2}
.
\end{equation}

\section{Identification of the energy scales}
\subsection{Characteristic energy of a disordered superconductor}
To take into account the large longitudinal dimensions of the junction,
we note that
convolutions of the exponential envelope of the transmission
coefficients lead to an
enhancement of $\lambda_\omega \epsilon^2$ by a factor $k_F^2 L_x L_y$
for extended interfaces. More precisely, combining diffusion
modes for $0$ to $x$ and from $x$ to $L$ leads to convolutions
of the form
\begin{equation}
\int_0^L \exp{(-x/\xi)} \exp{(-(L-x)/\xi)} dx= L \exp{(-L/\xi)}
,
\end{equation}
where the one dimensional case with two evanescent diffusion
modes is considered for simplicity.

We deduce from Eq.~(\ref{eq:lambda-om})
the characteristic energy $\hbar \omega_c$ of a dirty
superconductor:
\begin{equation}
\label{eq:omth}
\frac{\hbar\omega_c}{\epsilon_F}= \frac{1}{|\epsilon|}
\frac{1}{(k_F l_e)^{\alpha}}
\frac{1}{\sqrt{k_F^3 L L_x L_y}}
\end{equation}
The parameter 
of interactions $v$ in Eq.~(\ref{eq:omc0}) is
independent on $\Delta$,
as expected for a mechanism due to electron-hole conversion below the
gap \cite{BTK}; the lowest of the
characteristic energy $\hbar \omega_c$ and the voltage $e V_b$,
rather than $v \epsilon_F$, is
an indicator of the strength of weak localization-like quantum
interference effect in subgap transport.
The form of the transmission coefficients puts the
constraint $|\epsilon|<1$. The value $\epsilon=1$ favoring
EC with respect to CAR and minimizing $\hbar \omega_c$
in Eq.~(\ref{eq:omth}) for a static superconducting gap
is compatible with the previous picture of a fluctuating chemical
potential (see Fig.~\ref{fig:interweaving}),
and with the large EC crossed signal measured
experimentally at zero bias in Ref.~\onlinecite{Russo}.

\subsection{Comparison with experiments}
We use $l_e\simeq 2\,$nm from Ref.~\onlinecite{Russo},
$\epsilon_F\simeq 5.3\,$eV, 
$k_F \simeq 1\,$
\AA$^{-1}$ for Nb. In experiments, $\hbar \omega_c$ is
limited by
$l_\varphi\simeq0.1\div0.2\,\mu$m (as obtained from
the inelastic electron-electron scattering time
$\tau_{e-e}\simeq 1\,$ns \cite{Russo}),
instead of the sample dimensions
$L_x$ and $L_y$ in Eq.~(\ref{eq:omth}), leading to
$\hbar\omega_c \simeq 21\div10\,\mu$eV, $11\div 6\,\mu$eV and
$6\div 3\,\mu$eV for $L\simeq 15$,
$50$, $200\,$nm respectively. 
Russo {\it et al.} \cite{Russo}
find experimentally $\hbar\omega_c\simeq 270,$ $50\,\mu$eV, for $L\simeq 15$,
$50$nm, and $\hbar\omega_c$
below the resolution threshold for $L\simeq 200\,$nm.
The theoretical approximation
underestimates the experimental $\hbar \omega_c$ 
in the regime $L\agt\xi$ 
($L=15$, $50\,$nm and $\xi=10\div15\,$nm in experiments \cite{Russo}).
The energy scale $\Delta$ is associated to objects of size $\alt \xi$
formed by the accommodated weak localization-like loops for $L\simeq \xi$.
The energy scale $\hbar \omega_c$ is thus 
enhanced by a cross-over to 
perturbative effects of weak localization-like loops, with
$\hbar \omega_c$ reaching $\sim\Delta$ for $L\sim\xi$,
as in the absence of weak localization-like loops
\cite{Melin-Feinberg-PRB,Duhot-Melin}.
The cross-over is not captured by Eq.~(\ref{eq:omth})
with $L \gg \xi$.

\section{Conclusions}
In summary, intermediate states due to
double Andreev processes induced by
a weak localization-like quantum interference effect
in subgap transport are at the root of
a very slow attractive interaction between evanescent quasiparticles.
The interaction arises in an out-of-equilibrium situation
if quasiparticles are
forced to travel through the superconductor over distances
exceeding the coherence length $\xi$, which does not 
apply to usual Andreev
reflection limited by $\xi$. 
The phase of the mesoscopic superconductor fluctuates with a 
characteristic time
$2\pi/\omega_c$ if the bias voltage energy $e V_b$ is smaller than
$\hbar \omega_c$, because of the large retardation
$\hbar/e V_b$ in 
double Andreev processes due to weak localization-like loops,
as opposed to interactions
retarded over the Debye frequency for a BCS superconductor.
The model explains the following experimental facts: 
i) the existence of an energy scale $\hbar \omega_c$
within the superconducting gap, decaying to zero as the
superconductor thickness $L$ increases;
ii) a change of sign from EC for $e V_b<\hbar \omega_c$ to
CAR for $e V_b>\hbar \omega_c$, and a disappearance of the
crossed signal for $\hbar \omega_c$ larger than a few
$\hbar \omega_c$;
and iii) the coupling of weak localization-like quantum interference effect
to a magnetic field.
A challenging issue is to account for the interplay between
weak localization-like quantum interference effect discussed here and
the coupling to
the electromagnetic field \cite{Levy}.

\section*{Acknowledgments}
The authors acknowledge crucial discussion with B. Dou\c{c}ot,
D. Feinberg, M. Houzet, F. Pistolesi, J. Ranninger,
and thank H. Courtois, S. Florens and
K. Matho for useful remarks on the manuscript.

\end{document}